\documentclass[12pt,twocloumn]{article}
\usepackage[utf8]{inputenc}
\usepackage[numbers,sort&compress]{natbib}
\usepackage{subfigure}
\usepackage[version=3]{mhchem}
\usepackage{booktabs}	
\usepackage{graphicx}
	\graphicspath{{./}{./Figures/}} 
\usepackage{authblk}

\title{Spinodal decomposition stabilizes plastic flow in a nanocrystalline Cu-Ti alloy}
\author[1]{J. M. Rosalie}
\author[2]{O.Renk}   
\author[1,2]{R. Pippan}
\affil[1]{University of Leoben, Leoben, Austria.}
\affil[2]{Erich Schmid Institute, Austrian Academy of Sciences, Leoben, Austria.}

\begin{document}

\maketitle

\begin{abstract} 
A combination of high strength and reasonable ductility has been achieved in a copper-1.7\,at.\%titanium alloy deformed by high-pressure torsion. Grain refinement and a spinodal microstructure provided a hardness of $254\pm2$\,$H_V$, yield strength of 800\,MPa and elongation of 10\%.  The spinodal structure persisted during isothermal ageing, further increasing the yield strength to 890\,MPa while retaining an elongation of 7\%. This work demonstrates the potential for spinodal microstructures to overcome the difficulties in retaining ductility in ultra-fine grained or nanocrystalline alloys, especially upon post-deformation heating where strain softening normally results in brittle behavior.
\end{abstract}

\paragraph{Keywords} 
Titanium-bronze, Severe plastic deformation, Precipitation strengthening, High-pressure torsion.

\section{Introduction}
Ultra-fine grained (UFG) and nanocrystalline (NC) alloys can develop extremely high strength due to the confined slip length for dislocations \cite{Valiev2007,EdalatiHorita2011,LimChaudry2002,ValievEstrin2016}. However, the reduction of the grain volume also prevents storage of, and interaction between dislocations, resulting in poor work hardening. As a consequence plastic deformation becomes unstable even at small strains \cite{meyers2006, dao2007, Tong2009} resulting in  poor ductility \cite{SongPonge2005,WangWangPan2003, meyers2006, dao2007}. This can be exacerbated if UFG and NC materials are relaxed either thermally \cite{HuangHansen2006,wang2004,AlhamidiEdalati2014} or mechanically \cite{moser2006, rupert2012a,rupert2012b}. While the reduction of defect densities within the grains and at grain boundaries cause further strengthening, subsequent deformation and the generation of defects cause strain softening and early localization within shear bands and the behavior at the sample scale is  brittle.

Huge efforts have been made to retain ductility in nanomaterials by stabilizing plastic flow. Aside from the well-known example of nanotwinned metals \cite{ChenLuLu2011}, other attempts to retain work-hardening focused on the introduction of intragranular precipitates. Studies on materials with  grain sizes in the hundreds of nanometers have demonstrated intragranular precipitation can  improve ductility \cite{ChengZhao2007,ShanmugasundaramMurty2006,KuwabaraKurishita2006,wu2015}.

However, for grain sizes smaller than 100\,nm, intragranular precipitation is barely achievable due to grain boundary segregation of solute, either dynamically during severe plastic deformation or by diffusion during subsequent heat treatment \cite{BembalgePanigrahi2019, DevarajWang2019,ValievKazykhanov2019}. The escape of solute can be accompanied by the formation of intergranular precipitates (i.e. heterogenous nucleation of the phases). While stable grain boundary precipitates could limit  grain coarsening, their effect on the work-hardening rate and ductility is limited. In addition, they may even deteriorate the performance as intergranular cracks can form or grow more readily. Consequently, a different strategy is required for the finest grain sizes. 

Recent studies on systems forming Guinier-Preston zones (GPZ) \cite{LomakinCastillo2019} and spinodal microstructures \cite{TangGoto2017, HibinoWatanabe2017, WatanabeHibino2016,TangHirosawa2019} offer promise of a solution to this problem. As these decomposition processes occur homogenously and are weakly affected by the presence of defects such as grain boundaries, such features might be retained for the finest grain sizes and so permit better control of the work hardening behavior and ductility of NC metals. Unfortunately detailed studies focusing in this direction are rare, especially when considering the sub-100\,nm grain size regime. 

A Cu-Ti alloy was used herein to assess the effect of heat treatments on the microstructure and mechanical properties of a spinodal NC alloy. Cu-Ti alloys have been extensively studied \cite{soffa2004high, LaughlinCahn1975,datta1976structure, semboshi2014age, nagarjuna1997effect} due to their strong age-hardening response. Peak strengths of up to 1400\,MPa for extensively cold-worked and aged 5.4\,wt.,\% Ti (7.1\.at.\%) and 760 MPa for similarly treated 1.5\,wt.\% Ti (2.9\,at.\%) alloy can be reached due to a fine dispersion of \ce{Cu4Ti} plates, which gradually develops via spinodal decomposition \cite{ChenDuan2015,ZhuYan2016,NagarjunaBalasubramanian1999}. However, although the  limited solubility of Ti in Cu should permit the  generation of  an extremely fine grained microstructure upon high pressure torsion (HPT), there are to date few detailed studies of microstructure and mechanical properties in aged NC Cu-Ti. Hence, the Cu-Ti system offers a model system to study decomposition processes at the NC grain scale and their impact on mechanical properties.

\section{Experimental details}

Chemically homogeneous button ingots of composition Cu-1.7\,at.\%Ti (1.3\,wt.\%) were produced by repeatedly arc-melting Cu (Goodfellow, 99.95\%) and commercial purity Ti. These ingots were cold rolled to a thickness of $\sim$1.2\,mm, corresponding to a strain of about $\epsilon\sim 1.7$. Disks of 8\,mm  diameter were cut from the rolled sheet using electric discharge machining (EDM). The discs were encapsulated in Ar-filled  tubes, solution treated  (973\,K, 24\,h) and water quenched. The solution-treated (ST) discs were stored in a freezer at approximately $-10^\circ$C to prevent any natural ageing process and served as the basis for all subsequent investigations.

The disks were deformed at ambient temperature in a quasi-constrained HPT device \cite{vorhauer2008onset} at 0.2 rotations per minute under a nominal pressure of 7.8\,GPa. Radial hardness profiles (Vickers microindentation, 0.3 or 0.5\,Kg loads) showed that 10 rotations were sufficient to reach homogeneity throughout the disk, except for the very center part. The deformed disks were subjected to a series of isothermal (up to five days) and isochronal (30 min) annealing treatments up to 823\,K. Undeformed and conventionally deformed (compressed to $\sim50\%$ thickness reduction) disks were heat-treated in a similar manner to provide a comparison.

Changes of mechanical properties were tracked using microhardness measurements and microtensile test on selected samples. Microtensile specimens were cut by EDM from the HPT disks, with the gauge section being located at a radius of $\sim2$ mm, having a gauge length of 2.5 mm and a cross section of $\sim0.5 \times 0.5$ \,mm$^{2}$. The tensile specimens were tested using a miniaturized testing device (Kammrath \& Weiß, Germany) at a cross head speed of 2.5\,$\mu$m\,s$^{-1}$, corresponding to a nominal strain rate of $10^{-3}\,s^{-1}$. 
Strain was measured locally on the sample surface using optical images taken at a frequency of 2.5 Hz. Virtual strain gauges were then applied to the images using digital image correlation software (GOM Correlate). A connection between the testing rig and the camera allowed for accurate synchronization of the force signal with the elongation data from the images. Reduction in area of the respective samples and the fracture surfaces were analyzed using a Zeiss Leo 1525 field emission gun SEM.

The microstructural evolution  of the NC Cu-Ti samples was examined in the axial direction using a JEOL 2100 F Cs corrected transmission electron microscope (TEM), operating at 200\,kV. Foils were produced by dimple grinding and polished to perforation using a Gatan 691 precision ion polisher, equipped with liquid nitrogen cooling. Grain sizes were determined by applying the line intercept method to multiple bright-field STEM images. A minimum of 100 intercepts was used in all cases and the uncertainty estimated from the standard error for the number of intercepts. 

\section{Results}

\subsection{Ageing response}

The effect of high-pressure torsion on the ageing response was examined by comparing these samples to conventional materials with micron-scale grain sizes. The solution-treated alloy had a hardness of 95\,$H_V$ (see Table \ref{tab-hardness-delta}), which increased to 152$\pm$6\,$H_V$ after compressive deformation. In contrast, HPT deformed samples reached a saturated hardness\footnote{(i.e. the hardness effectively became independent of the radial position)} of 254$\pm$2\,$H_V$,  after 10 revolutions. 

\begin{figure*}[htbp]
	\begin{center}
\subfigure[\label{fig-hardness-ioschronal}]{\includegraphics[width=0.48\textwidth]{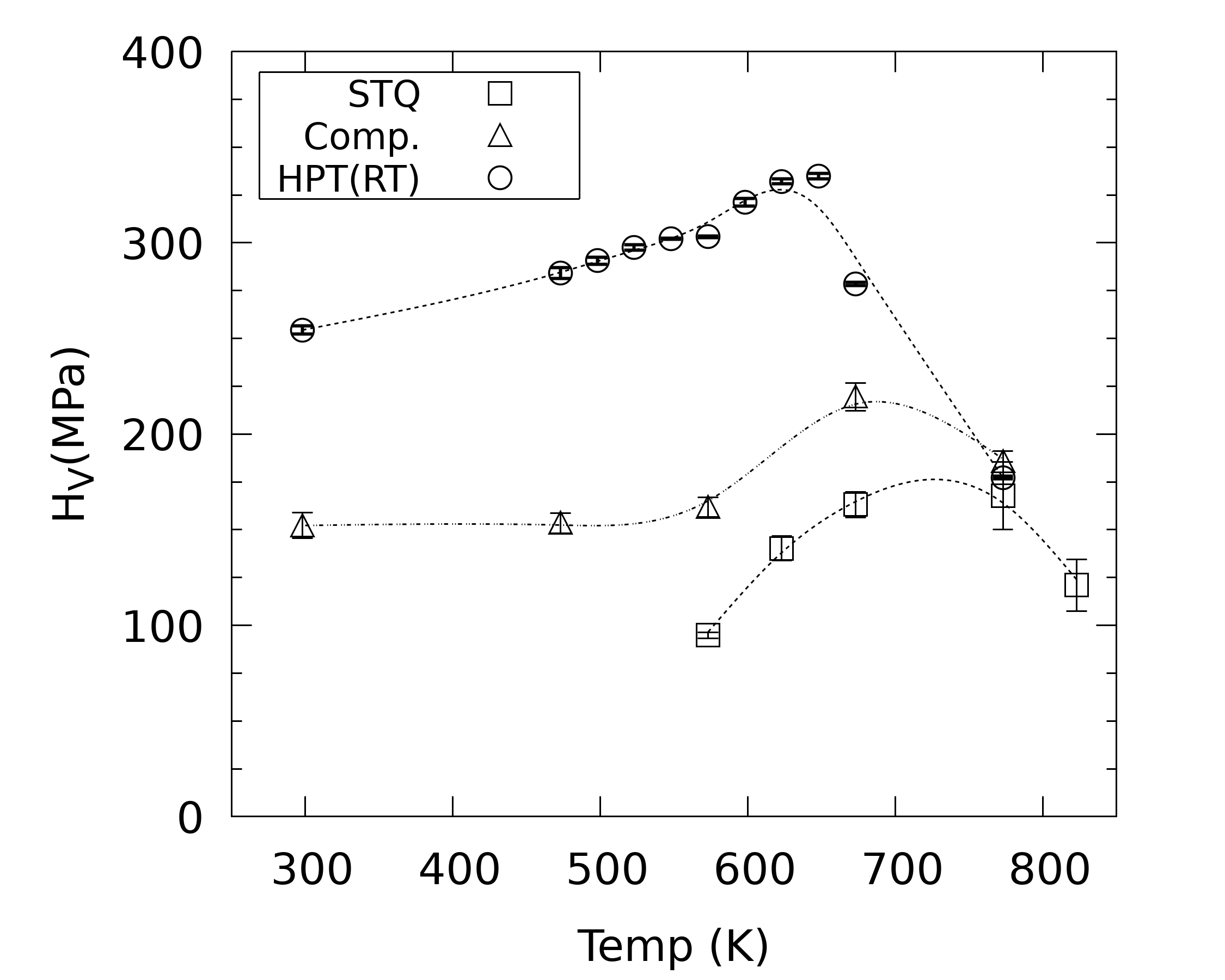}}
\hfill
\subfigure[\label{fig-hardness-iosthermal}]{\includegraphics[width=0.48\textwidth]{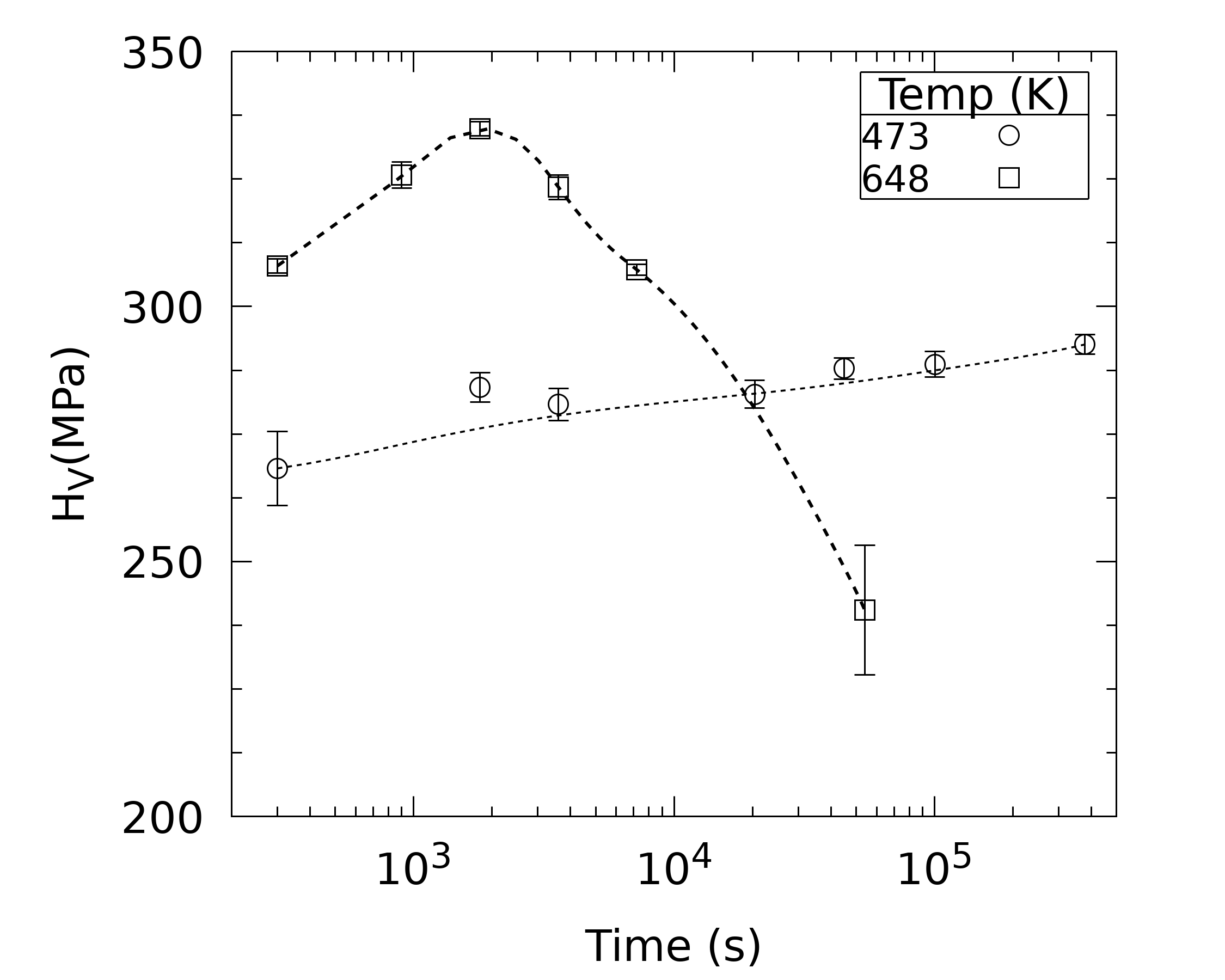}}
\caption{Hardness evolution of differently processed Cu-1.7\,at.\%Ti samples during a) isochronal ageing for 30 minutes
b) isothermal ageing of the HPT deformed samples at temperatures of 473 K and 648 K. Lines are include solely as a guide for the eye.}
\label{fig-vhn}
	\end{center}
\end{figure*}

Isochronal ageing curves, presented in Figure~\ref{fig-vhn}(a), show a strong ageing response in the STQ condition, which reached 1.68$\pm$20\,$H_V$ ($\Delta H_V$=73) for annealing at 773\,K. Compressive deformation increased the peak hardness to 233$\pm$17\,$H_V$ at a reduced temperature of 723\,K, but the ageing response was similar  ($\Delta H_V$=81) meaning the deformation caused little  net increase in extent of precipitation  hardening. 
A similar ageing-induced increase in hardness was seen in the HPT-deformed samples, but the higher as-deformed hardness resulted in a peak hardness of 303$\pm$0.01\,$H_V$ at 648\.K.  Temperatures higher than the optimum condition resulted in a dramatic softening, with the hardness of the HPT material approached the level of the ST and compressed specimens at 800\,K.

Isothermal ageing tests showed significant differences between ageing at 473\,K and 648\,K.  (see Figure~\ref{fig-hardness-iosthermal}) Typical age-hardnening behaviour was observed at 648\,K  with the hardness gradually declining to 2.3\,GPa after 8000\,s. In contrast, ageing at 473\,K resulted in a steady hardness plateau at 2.75\,GPa extending to at least 4$\times 10^5$\,s.

\begin{table*}[hbtp]
    \centering
        \caption{Comparison of hardness changes in each deformation condition. STQ: solution treated and quenched, STQC: STQ + compression, HPT: High-pressure torsion.
    \label{tab-hardness-delta}}
    \begin{tabular}{lllll}
        \toprule
        & \multicolumn{3}{c}{Hardness ($H_V$)}&
         Peak Temp.    \\ \cmidrule{2-4}
         State
         & Base
         & Peak
         & $\Delta  H_V$
         & K \\
         \midrule
         STQ & 
         90 &
         168 & 
         78 &
         723\\
         STQ-C & 
         153 & 
         233 & 
         80 &
         673\\
         HPT & 
         254 & 
         334 & 
         80&
         648\\
         \bottomrule
    \end{tabular}
\end{table*}

\subsection{Microstructure}
TEM characterization was performed to relate changes of hardness to the microstructural evolution, focusing especially on the response of the HPT deformed samples.

\subsubsection{Solution treated and quenched}
The  ST  material consisted of coarse ($\sim$50\,$\mu$m) grains (Fig.~\ref{fig-tem-overview-1}). This bright-field image shows two adjacent grains and the boundary separating them. The right-hand grain is in a two-beam condition, and dislocations and strain contrast can be noted. High-resolution transmission electron micrographs (HRTEM) (Fig. \ref{fig-tem-hrtem-1}) revealed the typical ``tweed'' contrast characteristic of  spinodal decomposition, resulting from local fluctuations in composition and lattice parameter  \cite{DattaSoffa1976,NagarjunaBalasubramanian1999,LaughlinCahn1975}. This is also evident in  satellite spots in Fast Fourier transforms (FFTs) (inset)\footnote{A Hann filter was used prior to applying the FFT, to remove high-frequency artefacts.}. 

\subsubsection{HPT-deformed and aged}

\begin{figure*}[hbtp]
\begin{center}
\rotatebox{90}{\begin{minipage}{0.4\textwidth}
\centering ST
\end{minipage}}
\subfigure[\label{fig-tem-overview-1}]{\includegraphics[width=0.4\textwidth]{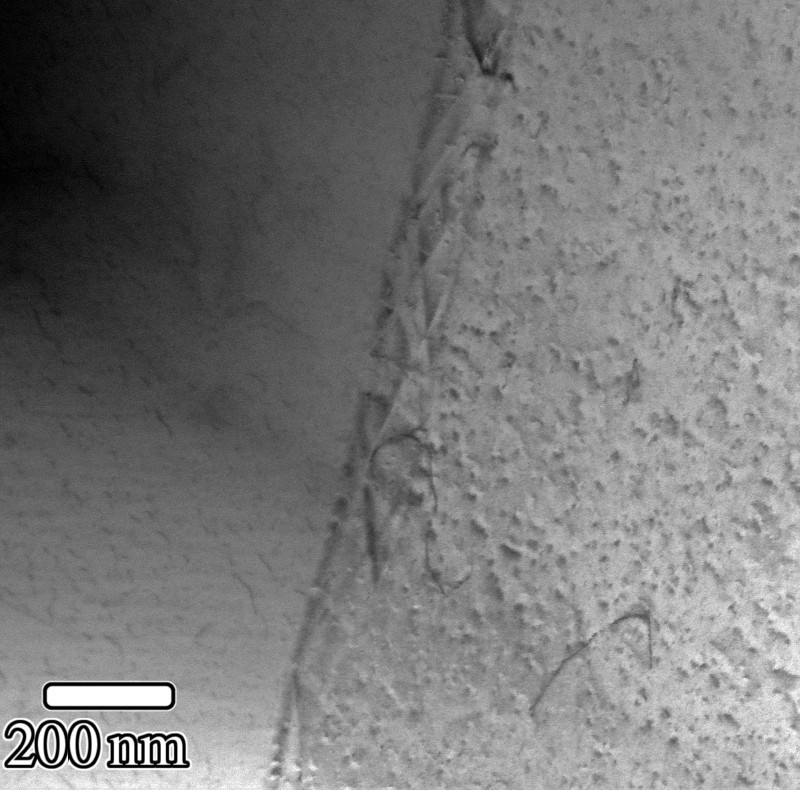}}\hfill
\subfigure[\label{fig-tem-hrtem-1}]{\includegraphics[width=0.40\textwidth]{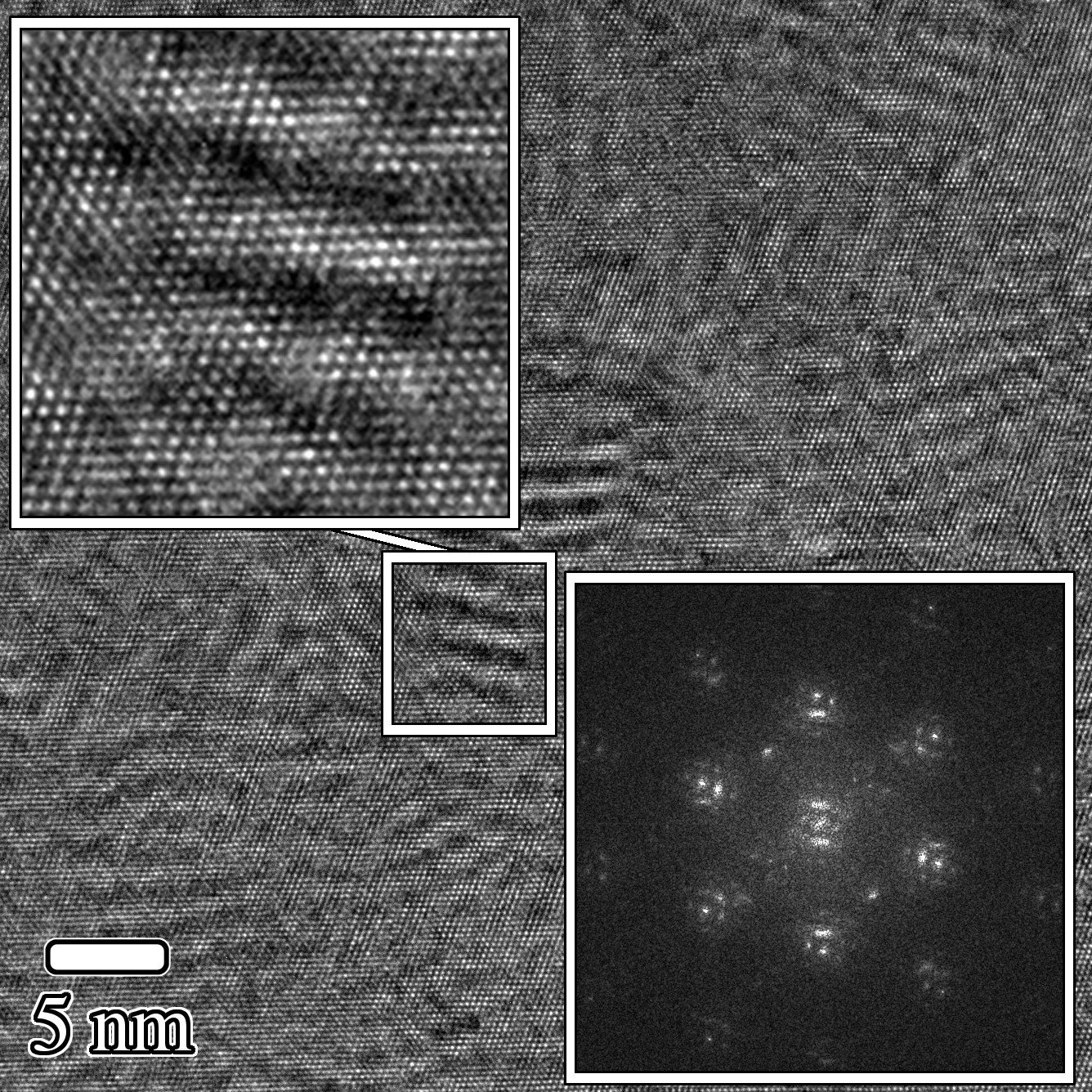}}

\rotatebox{90}{\begin{minipage}{0.4\textwidth}
\centering HPT
\end{minipage}}
\subfigure[\label{fig-tem-overview-2}]{\includegraphics[width=0.4\textwidth]{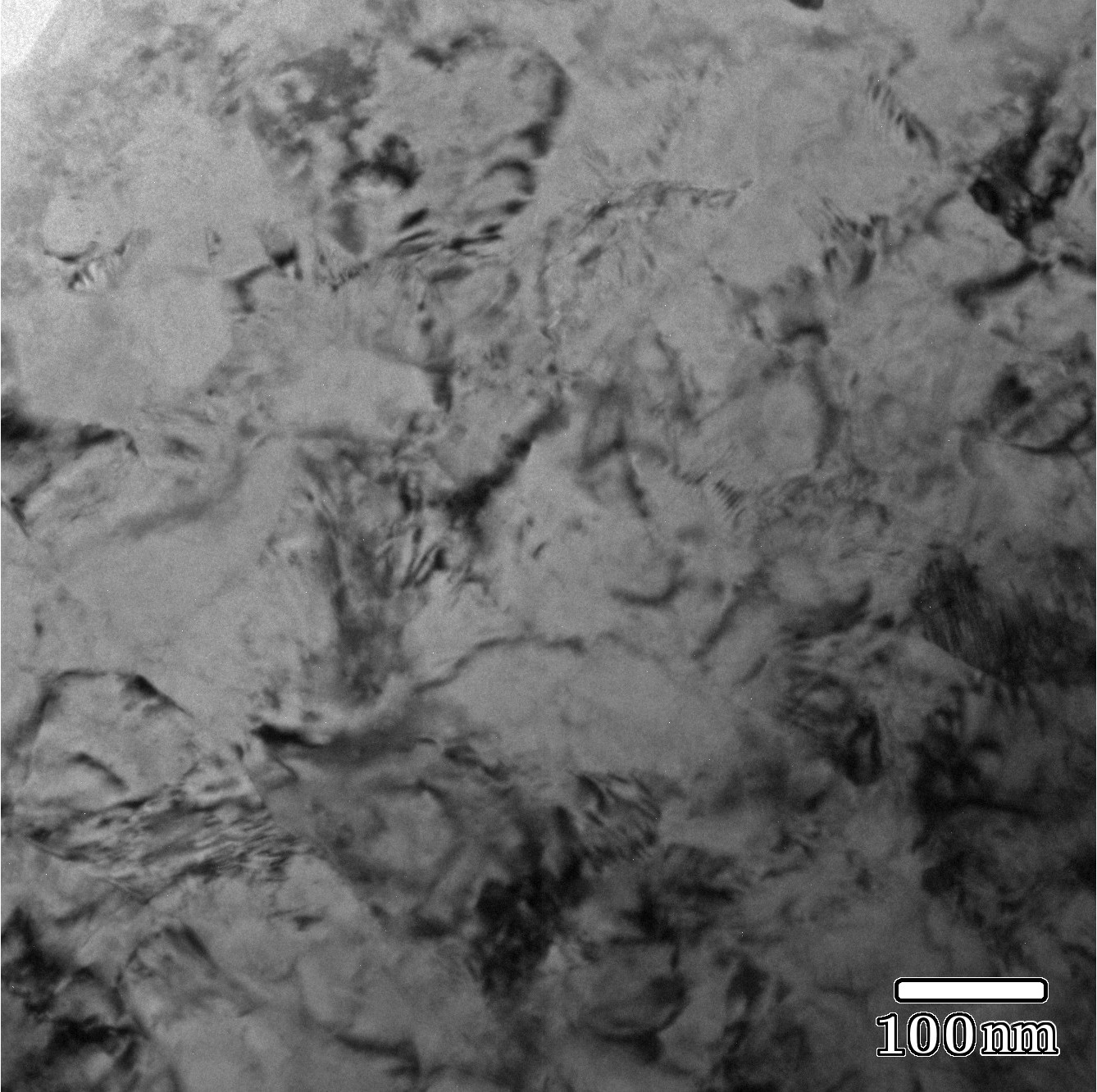}}
\hfill\,
\subfigure[\label{fig-tem-hrtem-2}]{\includegraphics[width=0.4\textwidth]{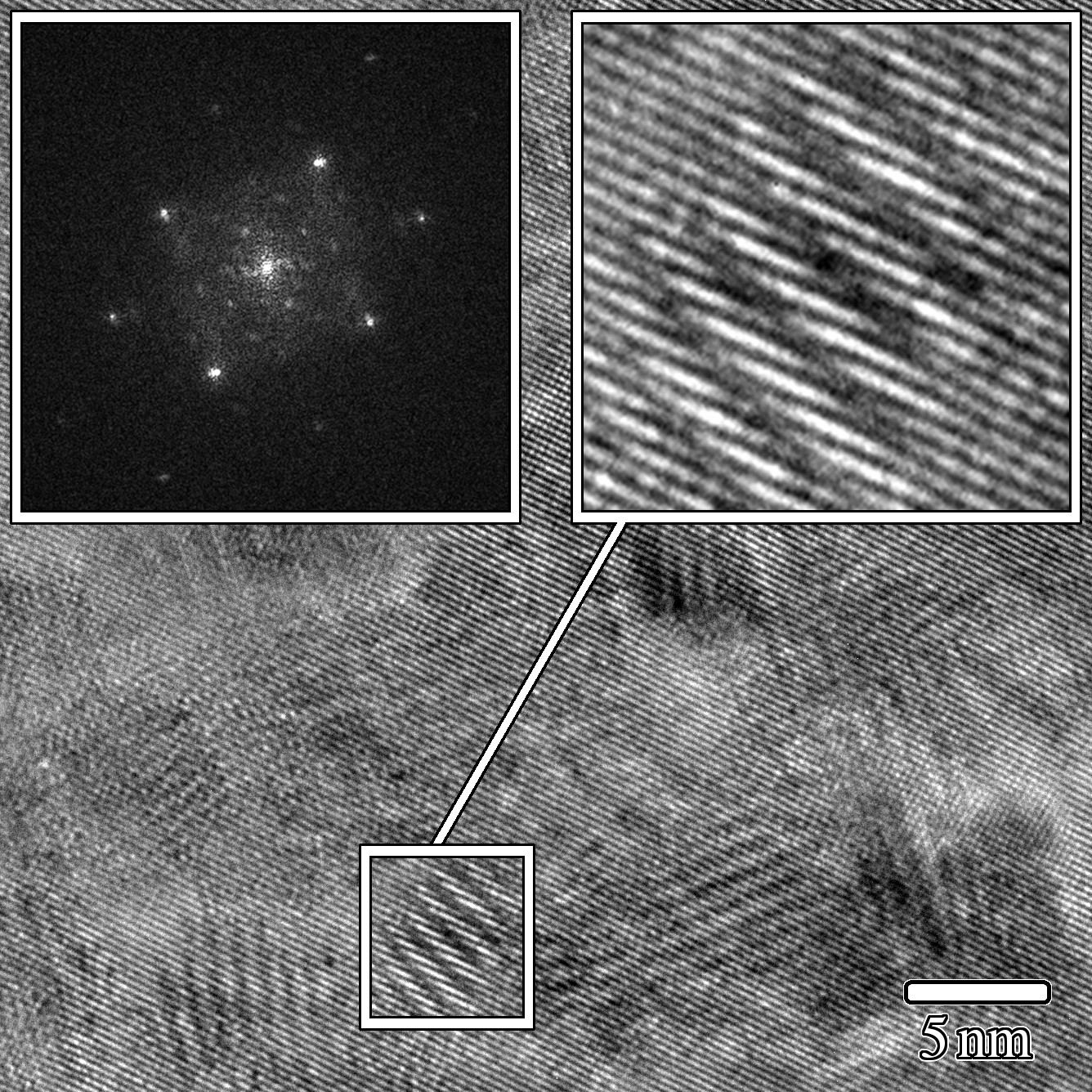}}

\rotatebox{90}{\begin{minipage}{0.4\textwidth}
\centering HPT+PA
\end{minipage}}
\subfigure[HPT+PA\label{fig-tem-overview-3}]{\includegraphics[width=0.4\textwidth]{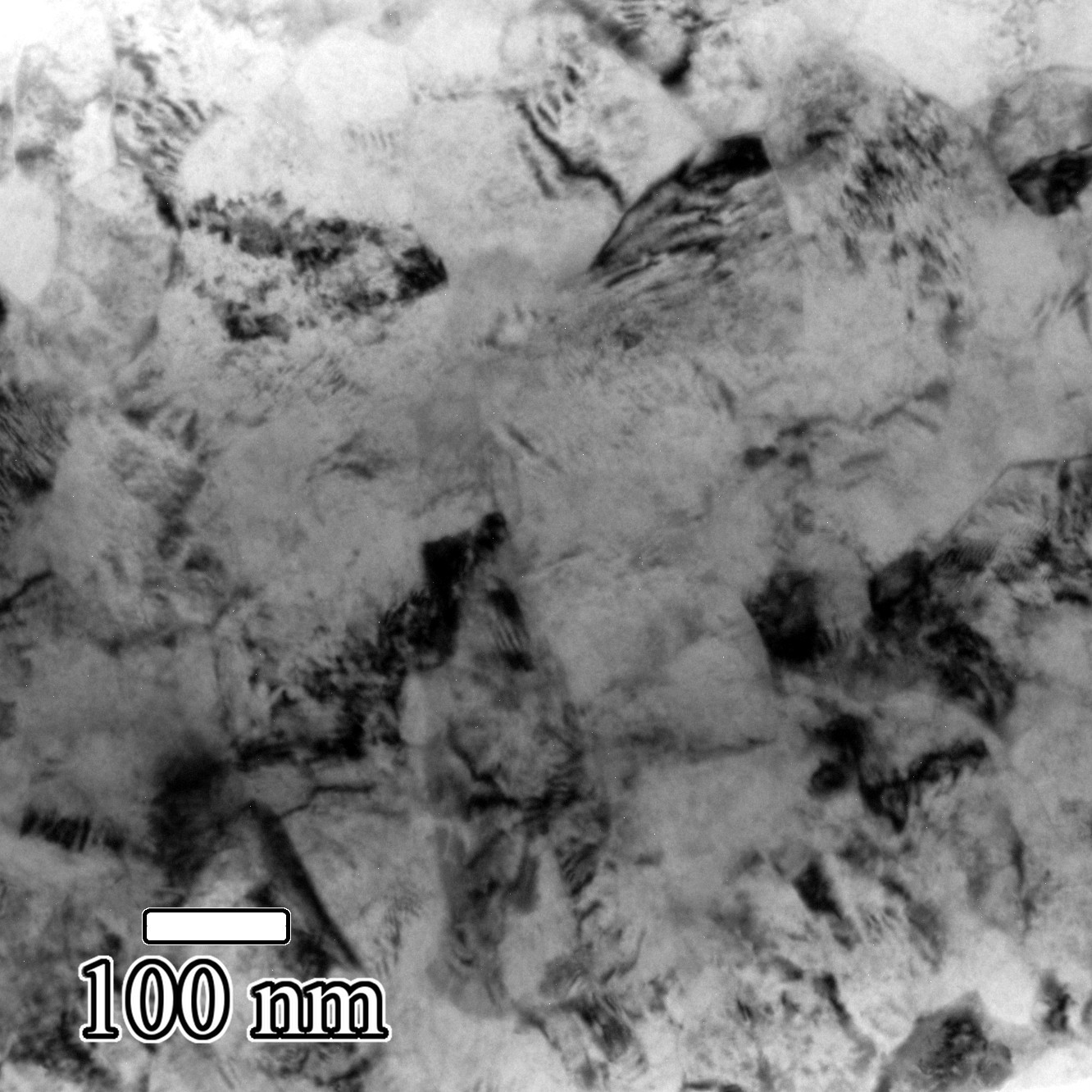}}
\hfill\,
\subfigure[HPT+PA\label{fig-tem-hrtem-3}]{\includegraphics[width=0.4\textwidth]{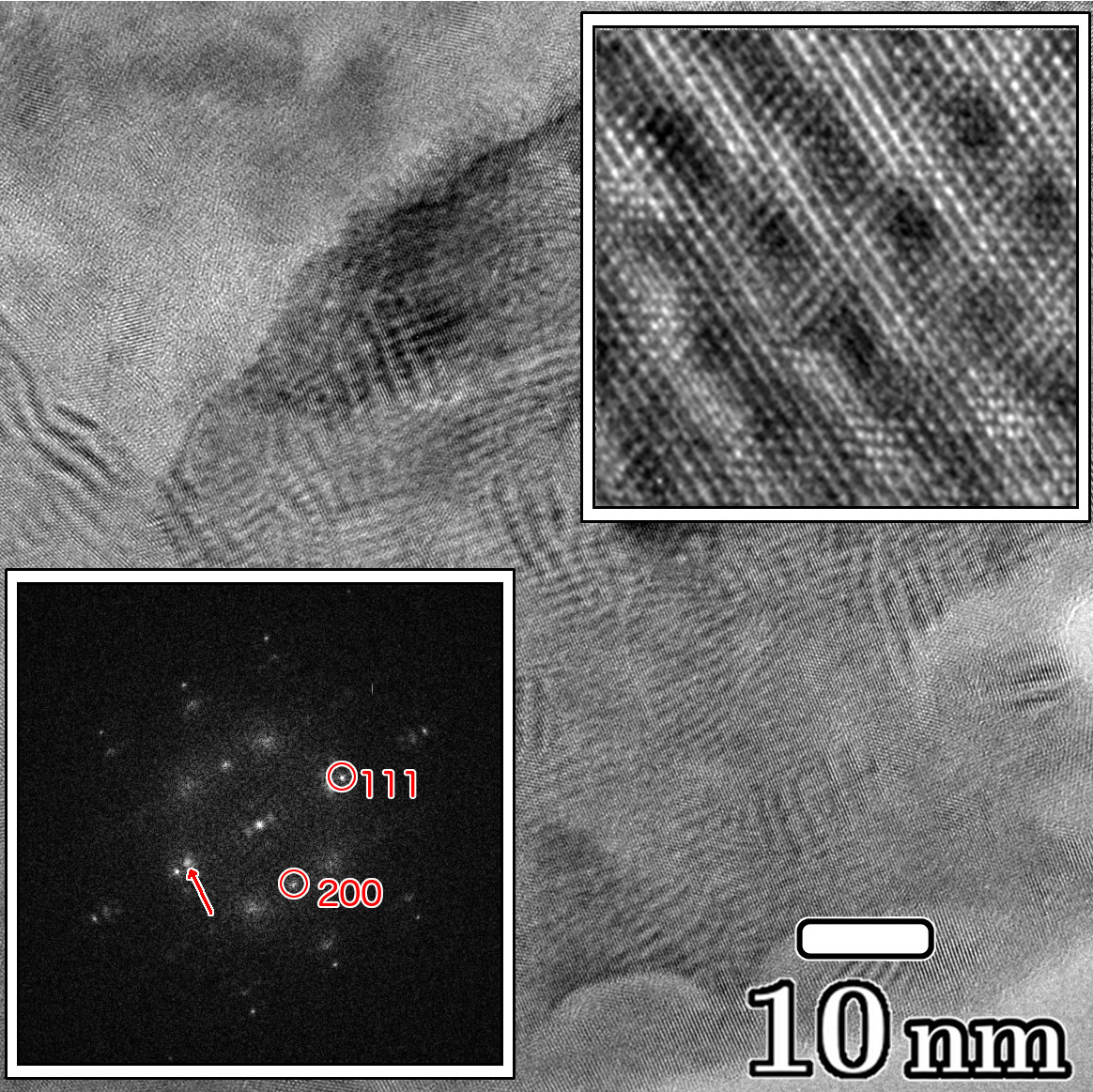}}

\caption{TEM images of the (a,b) ST, (c,d) HPT deformed and (e,f) peak-aged microstructures. \label{fig-tem-3}}
\end{center}
\end{figure*}

\begin{figure*}[htbp]
	\begin{center}
	\subfigure[\label{fig-modulations1}]{\includegraphics[width=0.48\textwidth]{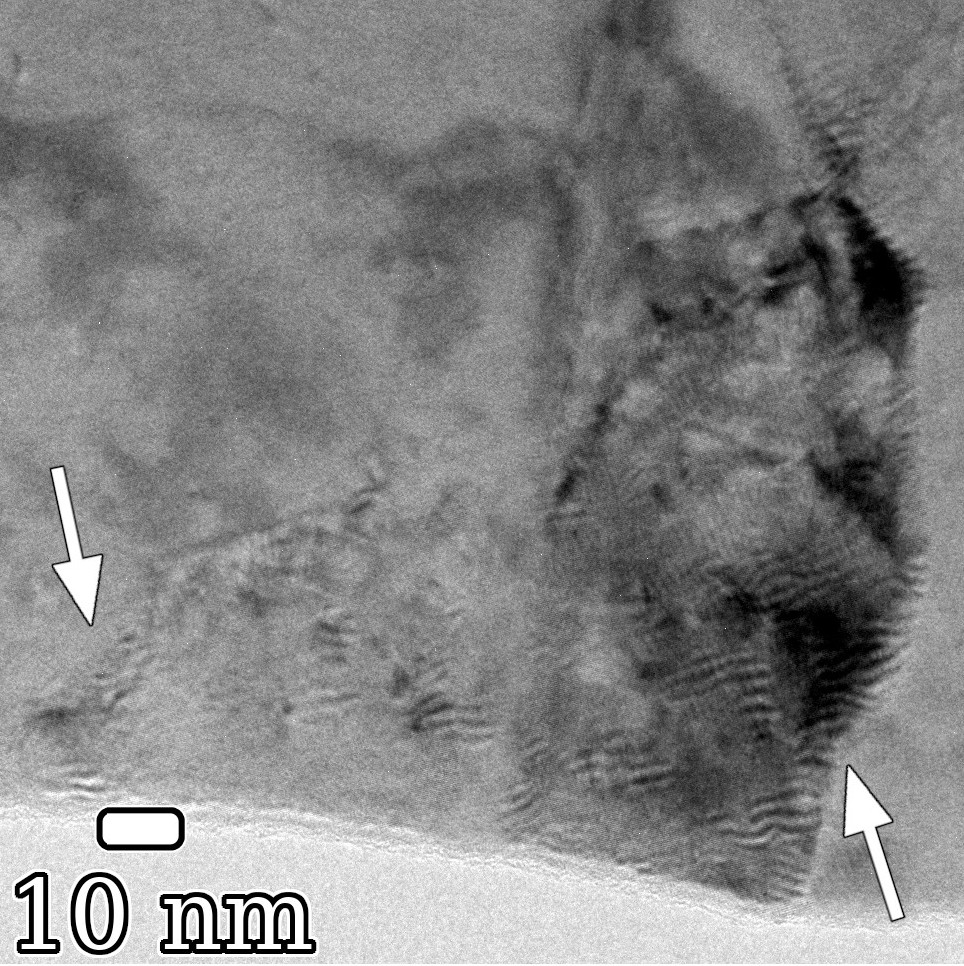}}
\hfill
	\subfigure[\label{fig-modulations2}]{\includegraphics[width=0.48\textwidth]{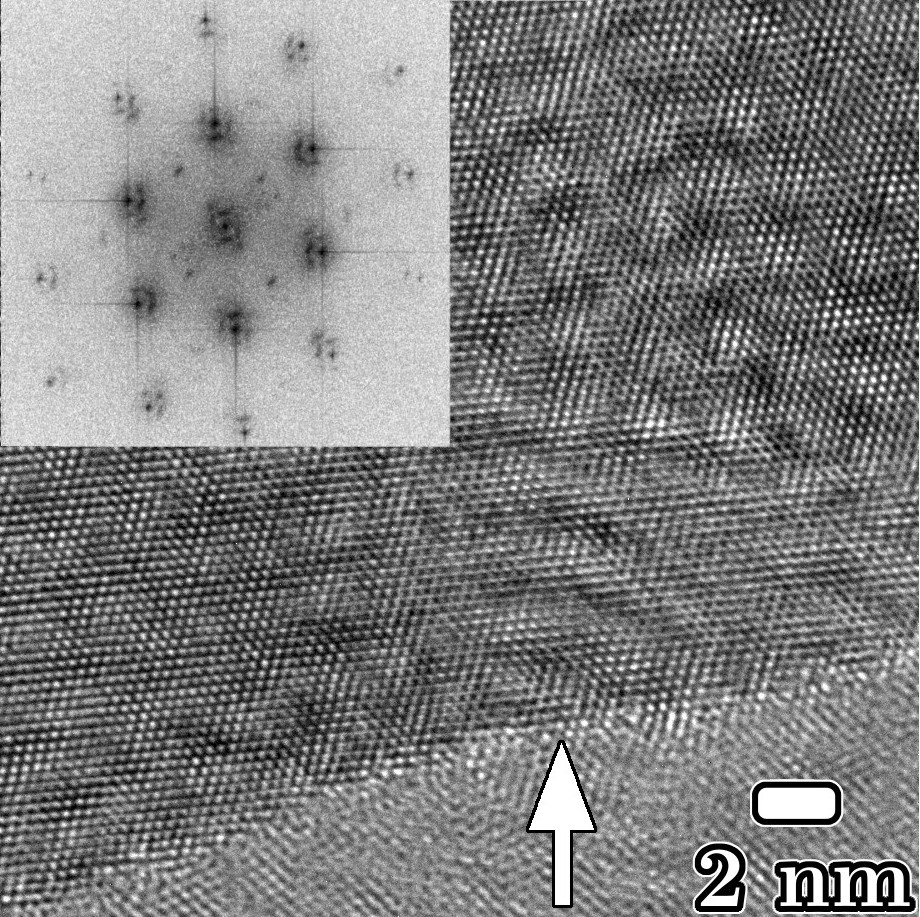}}
\caption{Modulated structure in Cu-Ti alloy HPT deformed and aged 30\,minutes at 573 K. (a) Diffraction contrast and (b) high-resolution TEM images show the modulated structure extending to the grain boundaries, as indicated by the arrows. \label{fig-modulations}}
	\end{center}
\end{figure*}

The spinodal structure is also present after HPT deformation (Fig. \ref{fig-tem-hrtem-2}) and even after subsequent ageing (Fig~\ref{fig-tem-hrtem-3}). However it appears to be broken up into several differently-oriented 5--10\,nm domains, which became more regular and well-defined after ageing. The modulated structure extends to the grain boundary, and grain boundary precipitation was seldom observed, compare Fig. \ref{fig-modulations}.

There was no indication of grain coarsening in the peak-aged condition, with the average grain size of $83\pm4$\,nm  effectively unchanged from the $82\pm5$\,nm measured for the as-deformed material. The grain boundaries in the peak-aged sample appear much sharper, indicative of defect relaxation. It should be noted, that the grain sizes measured in axial direction, can be considered only as an upper bound, while the minimum grain spacings (i.e. as seen in radial direction) will be significantly smaller.

\subsection{Tensile tests}
\begin{figure*}[htbp]
	\begin{center}
	\subfigure[Engineering\label{fig-tensile-eng}]{\includegraphics[width=0.48\textwidth]{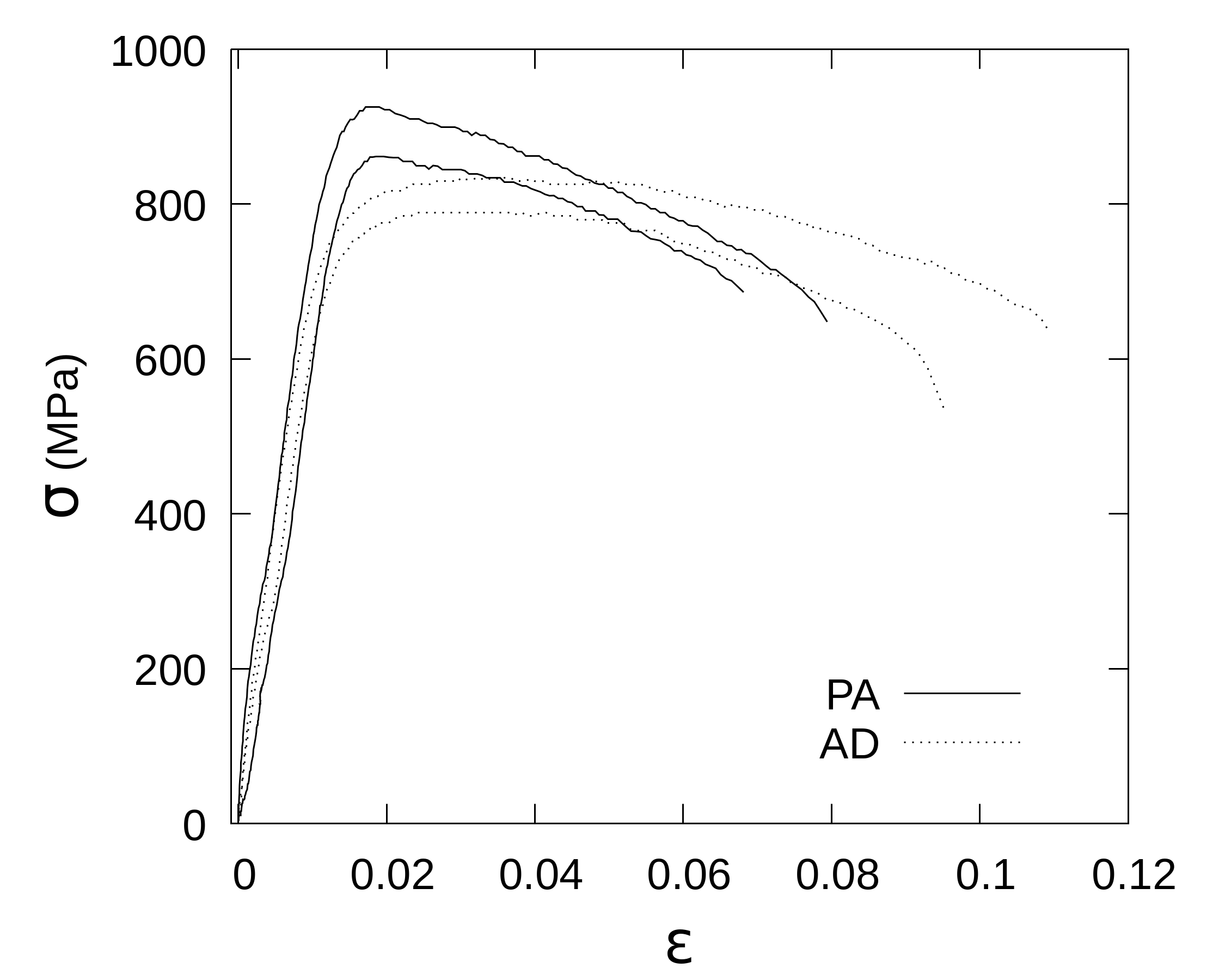}}
\hfill
	\subfigure[True\label{fig-tensile-true}]{\includegraphics[width=0.48\textwidth]{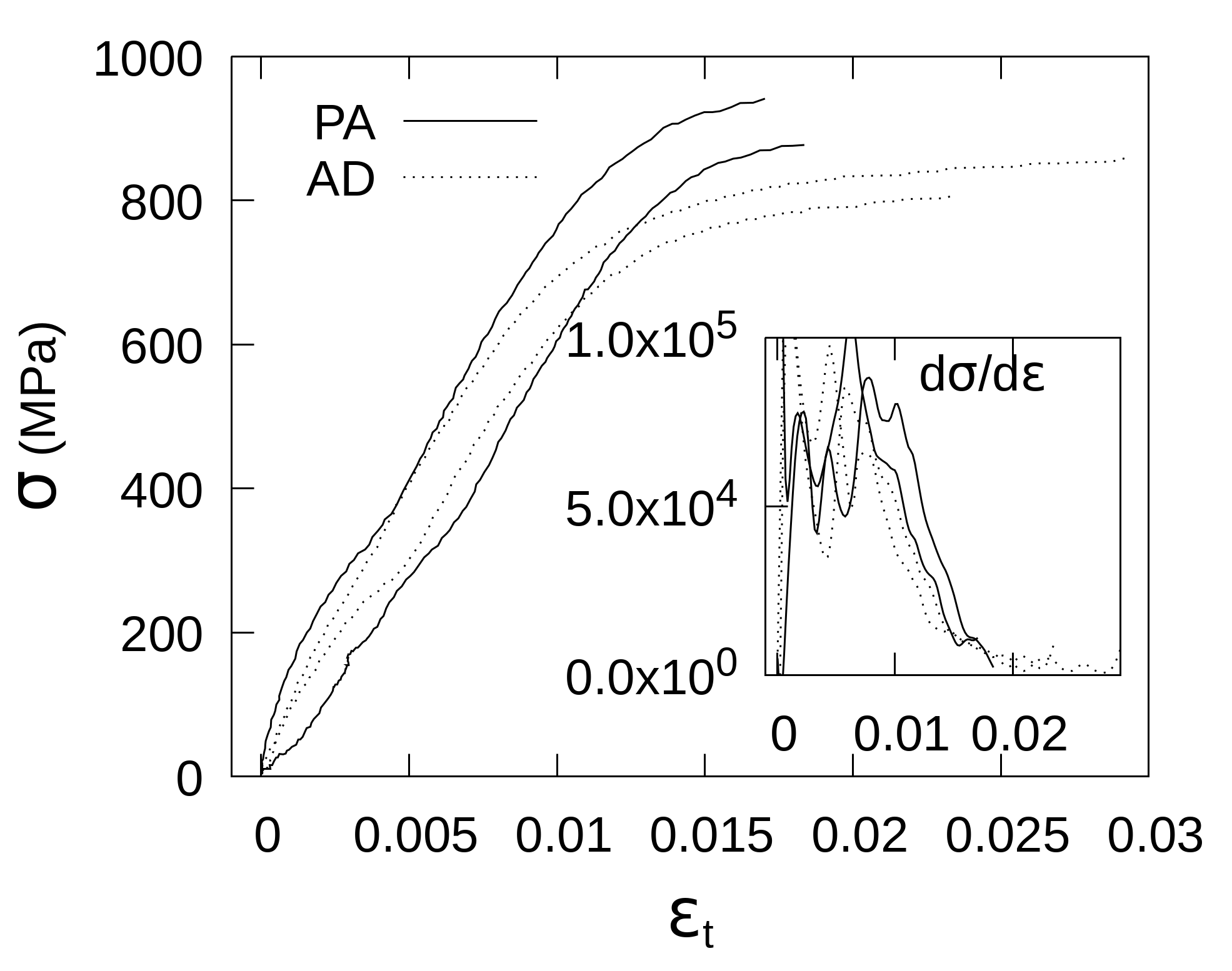}}
\caption{(a) Engineering and (b) true stress-strain curves for the as-deformed  (AD) and peak-aged (PA)  conditions (see Table~\ref{tab-tensile}). The inset in (b) shows the work hardening rate. \label{fig-tensile}}
	\end{center}
\end{figure*}

Microtensile tests on the as-HPT deformed and peak aged samples showed the high strength typical of severely deformed metals and alloys. Representative engineering and true stress-strain curves of the tensile tests are shown in Fig.~\ref{fig-tensile-eng} and \ref{fig-tensile-true}, respectively. A tensile strength of $\sim$800\,MPa was obtained for the deformed material, increasing to 890\,MPa at peak age. Despite this, both materials remained ductile, with an elongations to failure of 9\,\% and 6\,\%, respectively. Work hardening rate curves (inset in Fig~\ref{fig-tensile-true}) also indicate appreciable hardening to true strains $>$0.02, being slightly increased in case of the peak aged samples. The higher work hardening rate in case of the peak aged samples is further evident when considering the true fracture stress of the samples, compare Table \ref{tab-tensile}. In case of the PA samples the average true fracture stress is about 1300 MPa, significantly higher than the one calculated for the as-HPT deformed specimens. But apart from the higher absolute values, also the relative hardening capacity (i.e. the stress increment form the yield stress to the true fracture stress) is clearly enhanced for the PA samples.

\begin{table*}[hbtp]
    \centering
        \caption{Mechanical properties of the HPT deformed and additionally aged Cu-1.7\,at.\%Ti samples in tension.\label{tab-tensile}}
    \begin{tabular}{llllll}
        \toprule
         State
         & YS
         & UTS
         & Elong.
         & Area. redn. 
         & true frac. stress\\
         &  \multicolumn{2}{c}{(MPa.)}
         & \multicolumn{2}{c}{(\%)}
         & {(MPa)}\\
         \midrule
         HPT &
         700&
         810$\pm$40&
         9$\pm$1 & 43.1 $\pm$7.5 & 1035 $\pm$53
         \\
         HPT-PA &
         800&
         890$\pm$60&
         6$\pm$1 & 48.5$\pm$4.7 & 1309 $\pm$157
         \\
         \bottomrule
    \end{tabular}
\end{table*}

\subsubsection{Post-fracture examination}

\begin{figure*}[htbp]
	\begin{center}
	\includegraphics[width=0.9\textwidth]{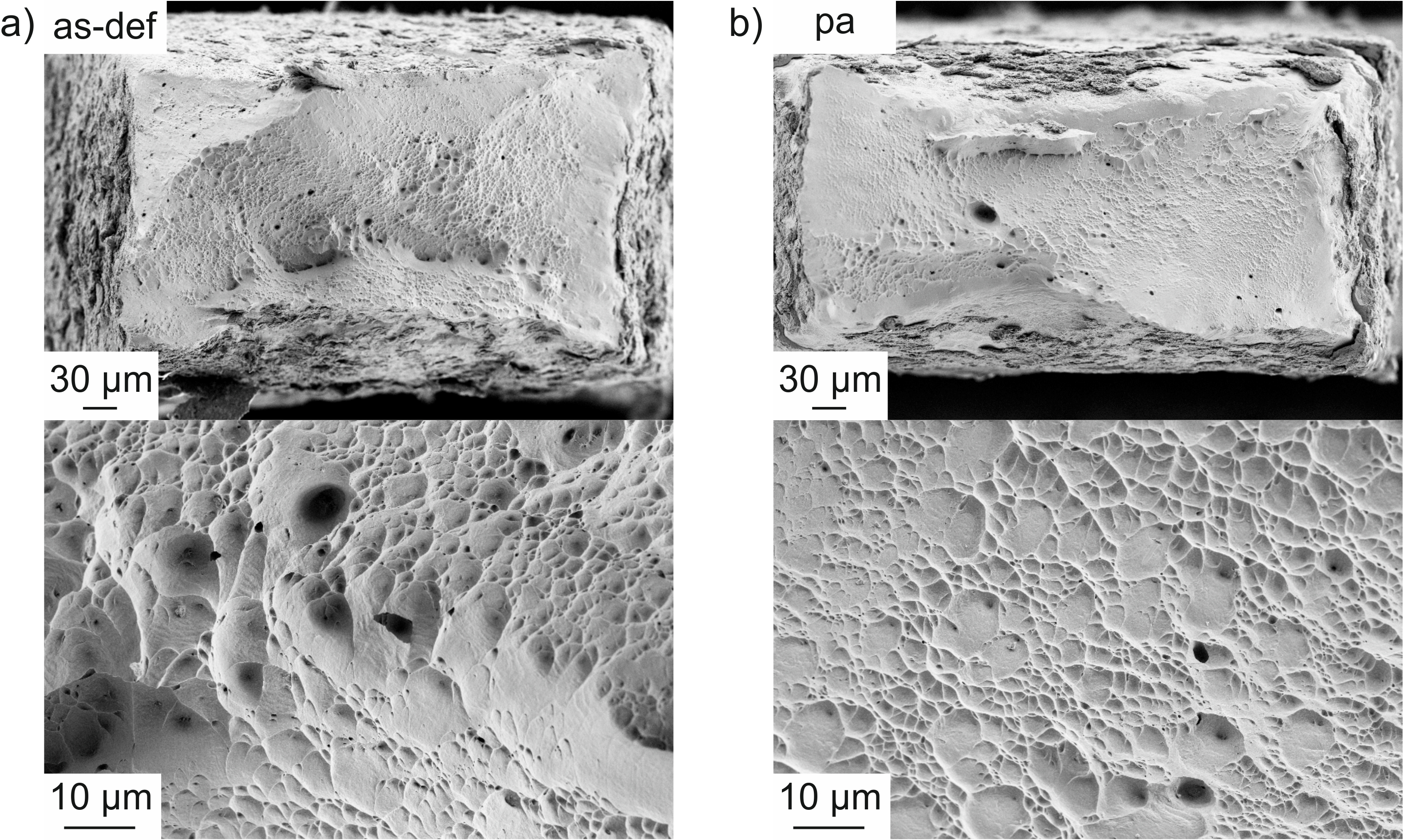}
\caption{SEM images of the fracture surfaces of the (a) as-deformed and (b) peak-aged Cu-Ti tensile samples.\label{fig-fracture}}
	\end{center}
\end{figure*}

SEM examination of the fracture surfaces of the tensile sample provided additional evidence of substantial ductility at a micrometer scale  (Figure~\ref{fig-fracture}). The low magnifcation images (above) show a notable area reduction (see Table~\ref{tab-tensile}), with extensive microdimpling visible at higher magnification. These microductile features are commonly observed in UFG or NC metals. However, the dimple size observed here is significantly larger than the grain size, while for UFG or NC materials the dimple size is typically only a few times the grain size \cite{hasnaoui2003dimples,kumar2003deformation}. The significantly larger dimple size observed here, is further indicative of a pronounced intragranular dislocation activity. Roughly spheroidal pores may indicate pull out of undissolved Ti particles.

Despite the fracture surface of the peak-aged sample being flatter and more uniform than that of the as-deformed material, the images show no evidence of intragranular fracture.

\section{Discussion}
This work has demonstrated that NC Cu-Ti alloys have retain moderate ductility, even when subjected to isothermal ageing. Moreover, the NC alloy displayed exceptionally high strength for a low Ti content. The hardness of 334\,$H_V$ for the PA condition compares favourably with literature reports of Cu with a grain size of 30--40\,nm \cite{WangWangPan2003}, with a much higher elcongation of 7\%. Although grain sizes being by a factor of two larger were measured in axial direction, the minimum grain spacing in radial direction may reasonably compare to these values. The Cu-Ti alloy also show good thermal stability, as shown in  Fig~\ref{fig-hardness-iosthermal}. The factors underlying this promising combination of properties are discussed in the following section, by comparing the microstructure and mechanical properties with the coarse-grained material.

\subsection{Coarse-grained Cu-Ti}
The solution treated material was comprised of a modulated structure formed by spinodal decomposition of the alloy during quenching (Fig.~\ref{fig-tem-hrtem-2}), as consistent with previous report on this system \cite{LaughlinCahn1975}, albeit at lower solute contents. This microstructure already offered significant strengthening, as shown by a hardness of 0.90\,GPa, compared to approximately 0.50--0.55\,GPa for well-annealed Cu of similar grain size \cite{LimChaudry2002}.

The ageing response of the ST and ST-and-compressed samples can be attributed to decomposition and precipitation processes, as expected for this alloy \cite{DattaSoffa1976, SoffaLaughlin2004, SukHong2009, SemboshiNishida2011}. During the early stages of ageing this is expected to take place via an increase in the amplitude of the composition fluctuations, followed by the precipitation of \ce{Cu4Ti} precipitates. 

In conventional, coarse-grained CuTi alloys the loss of mechanical properties on overageing is attributed to the discontinuous precipitation of intermetallic phases on the grain boundaries. The redistribution of solute to these phases results in the loss of the intragrannular strengthening structures (spinodal modulations and/or fine Cu$_4$Ti precipitates).

\subsection{Nanocrystalline Cu-Ti}
HPT deformation resulted in extensive grain refinement, and a hardness of 2.54\,GPa. This was almost twice the value of HPT deformed pure copper \cite{renk2014} and a five-fold increase on well-annealed pure Cu \cite{LimChaudry2002}. HPT deformation further accelerated the ageing kinetics, reducing the isochronal peak-ageing temperature to 648\,K, compared to 723\,K for ST and 673\,K for compressed material.
This is consistent with detailed studies on the age hardening characteristics of Cu-Ti alloys, indicating a shift of the peak ageing temperature to lower values with increasing pre-deformation \cite{nagarjuna1997effect, NagarjunaBalasubramanian1999}.

It was surprising that a similar modular structure was present after HPT deformation to equivalent strain of $\sim136$ (at a disk radius $r=3\,$mm) (Fig.~\ref{fig-tem-hrtem-3}), and it will be made clear in this section that this influences the mechanical properties in a similar way as for the coarse-grained material. In fact, the Cu-Ti system showed no evidence of a perfect single phase supersaturated solid solution after HPT, in contrast with numerous other alloy systems, including many which are immiscible under normal conditions \cite{PouryazdanSchwen2012,BachmaierKerber2012,GuoRosalie2017,MaSheng2000}. This could be either related to an enhanced diffusion during HPT or fast relaxation at room temperature handling enabled by the large defect densities present. While the first scenario seems plausible based on reports claiming superfast diffusion (e.g. \cite{WangLu2010}, where the self-diffusion of Cu increased by $10^{19}$  \cite[Refs 4,14 in]{Birringer1989}), a fast natural ageing (i.e. decomposition) allowing the material to rapidly relax to the metastable modulated structure seems more plausible. If decomposition would occur already during processing, the modulated structure should appear kinked or deformed at least in certain grains. Moreover, fast natural ageing was already reported for this kind of alloys within reasonably short time frames, e.g. Refs. \cite{WatanabeHibino2016,HibinoWatanabe2017}. Although difficult to prove unambiguously, a supersaturated solid solution may thus be present directly after HPT deformation.

A fast relaxation might further explain why, in contrast to the starting microstructure the typical\ ``tweed'' contrast of the decomposed structure is composed of multiple nanoscale ``domains'' as can be seen in the  HR-TEM micrographs (Fig~\ref{fig-tem-hrtem-3}).  Spinodal decomposition in this system involves composition fluctuations in the (soft) $\langle 100 \rangle$ directions \cite{LaughlinCahn1975} hence three variants are possible. Such nanodomain formation would be consistent with the rapid relaxation of the structure at very short length scales.

Relaxation of intra- and intergranular defects during annealing of UFG and NC metals (``hardening by annealing'') can be a significant contributor to strength \cite{HuangHansen2006,AlhamidiEdalati2014, renk2015, rupert2012a,rupert2012b}. For nanomaterials, relaxation of grain boundaries and intragranular dislocation densities, potentially accompanied by solute segregation, is known to enhance the flow stress of the material as dislocation sources need to be activated to ensure plastic flow \cite{hasnaoui2002non, HuangHansen2006}. This phenomenon is reported even in modulated systems such as Al-Zn \cite{AlhamidiEdalati2018} and manifests as a  hardness plateau even after prolonged annealing times as long as grain growth is absent \cite{rupert2012b, renk2015,renk2015hardening}. 
However, although some relaxation of the grain boundaries appears to have occurred (compare Figs. \ref{fig-tem-hrtem-2} and \ref{fig-tem-hrtem-3})
the isothermal ageing curves  (Fig~\ref{fig-hardness-iosthermal}) show a substantial decline in hardness after extended ageing, suggesting that relaxation strengthening is not the sole contribution to the age-hardening of the material. The domain-like structure is retrained even upon heating, as shown in Figure~\ref{fig-tem-overview-3}, \ref{fig-tem-hrtem-3} and \ref{fig-modulations}. The amplitude of the composition fluctuations would be expected to increase during spinodal composition. This effect, in addition to  defect relaxation and/or segregation results in increased strengthening during ageing \cite{Cahn1963,Dahlgren1977}. 

The overaging mechanism here is expected to be similar to that occur in the CG case, and preliminary observations on overaged NC samples did detect grain boundary phases. In this case the loss of strengthening is probably exacerbated by grain coarsening; although grain boundary particles can act as pining sites, rapid, preferential diffusion along the grain boundary could rapid coarsen the larger particles, at the expense of the loss of the smaller ones, and hence reduce the number of pining sites. 

To quantify the contribution of modulation strengthening vs. hardening by annealing to the 30\% strength increase would require careful consideration of the models describing the strength increase by spinodal decomposition (e.g.\cite{Cahn1963,Dahlgren1977}), and other contributions such as grain-size strengthening. This would be best suited to techniques in which \textit{in situ} measurements can be made during ageing, for example small angle x-ray scattering.

From the results it is evident that it is the decomposed structure, not the minimum obstacle spacing which governs strength. The large fraction of grain boundaries present in the NC sample still plays a dominant role. Although the wavelength (i.e. the modulation period) of the decomposed structure is more or less independent of the grain size or the treatment prior to annealing (see Fig.~\ref{fig-tem-3}), for the HPT deformed NC samples, the PA is about 1.2 GPa larger than for the coarse-grained PA samples, Table \ref{tab-hardness-delta}. This suggests, that even for nanoscaled coherent or semi-coherent decomposed or precipitated structures a NC grain structure can add significant additional strength. Moreover, apart for further strengthening the NC grain structure could help to retain ductility for such systems, as discussed in the remainder.

However, from the tensile tests it is evident, that no matter if defect relaxation occurs, the results differ significantly to systems without decomposition or precipitation. Anneal-hardening is generally associated with a massive reduction in ductility, as any defect (dislocation) introduced to the crystal after the heat treatment can readily propagate, resulting in a significant strain softening, shear band formation and early failure, compare Refs. \cite{wang2004effects, HuangHansen2006}. The peak aged samples studied here behave differently, with reasonable ductility and no evidence of macroscopic strain localization despite the gain in strength. This indicates that the decomposed structure is suitable to hinder dislocation motion and induce some work hardening.
This is well-known from coarse grained materials. Spinodally decomposed systems or clusters act as obstacles for dislocation motion, but as they are penetrable or shearable, the resistance against dislocation motion remains relatively weak. Hence, their strengthening effect to coarse grained metals generally comes along with sharp reduction of ductility \cite{NagarjunaSarma1999}, the opposite to what is observed here. Therefore, the observed larger work hardening in case of the peak aged samples and the almost unaffected ductility despite the higher strength explained by the increased amplitude of the decomposed regions seems surprising. As the decomposed structure hardly changes due to the grain size reduction (compare Fig. \ref{fig-tem-3}), the different effect on ductility might be rationalized based on the grain size. 

In case of the HPT deformed samples, the overall tensile strain is only on the order of up to 10 \%. Hence, the number of number of dislocations per grain involved is fairly limited and can be estimated by $0.1=N*b/d$. For NC structures and fairly low strains, the number of dislocations required thus not exceed several tens of dislocations. This may not be sufficient to shear decomposed structures or precipitates, hence, soft pathways may not be established, supported by the measured larger work hardening rates in case of the peak aged samples.
 
\begin{figure*}[htbp]
	\begin{center}
\subfigure[Coarse-grained spinodal\label{fig-work-hardening-1}]{\includegraphics[width=0.48\textwidth]{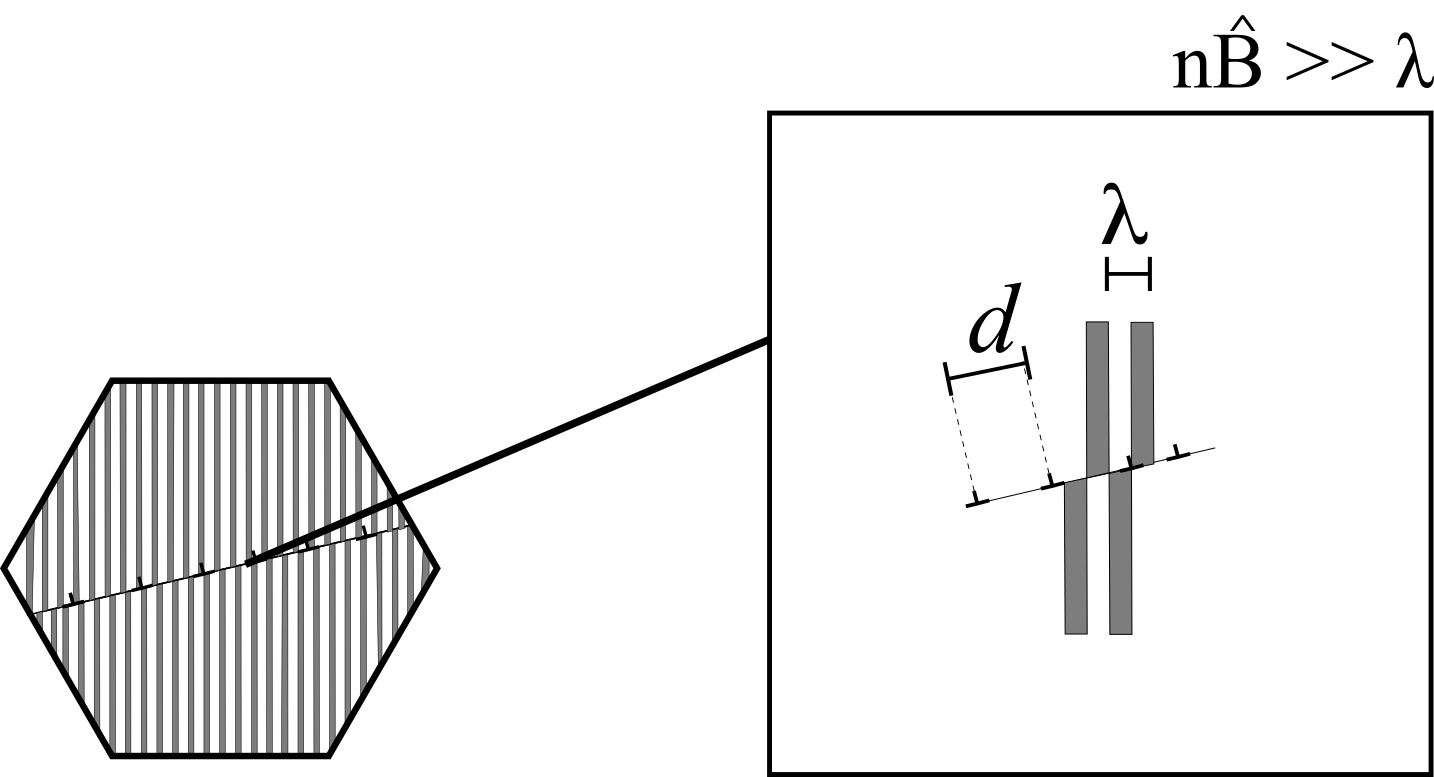}}
\hfill
\subfigure[UFG spinodal\label{fig-work-hardening-2}]{\includegraphics[width=0.48\textwidth]{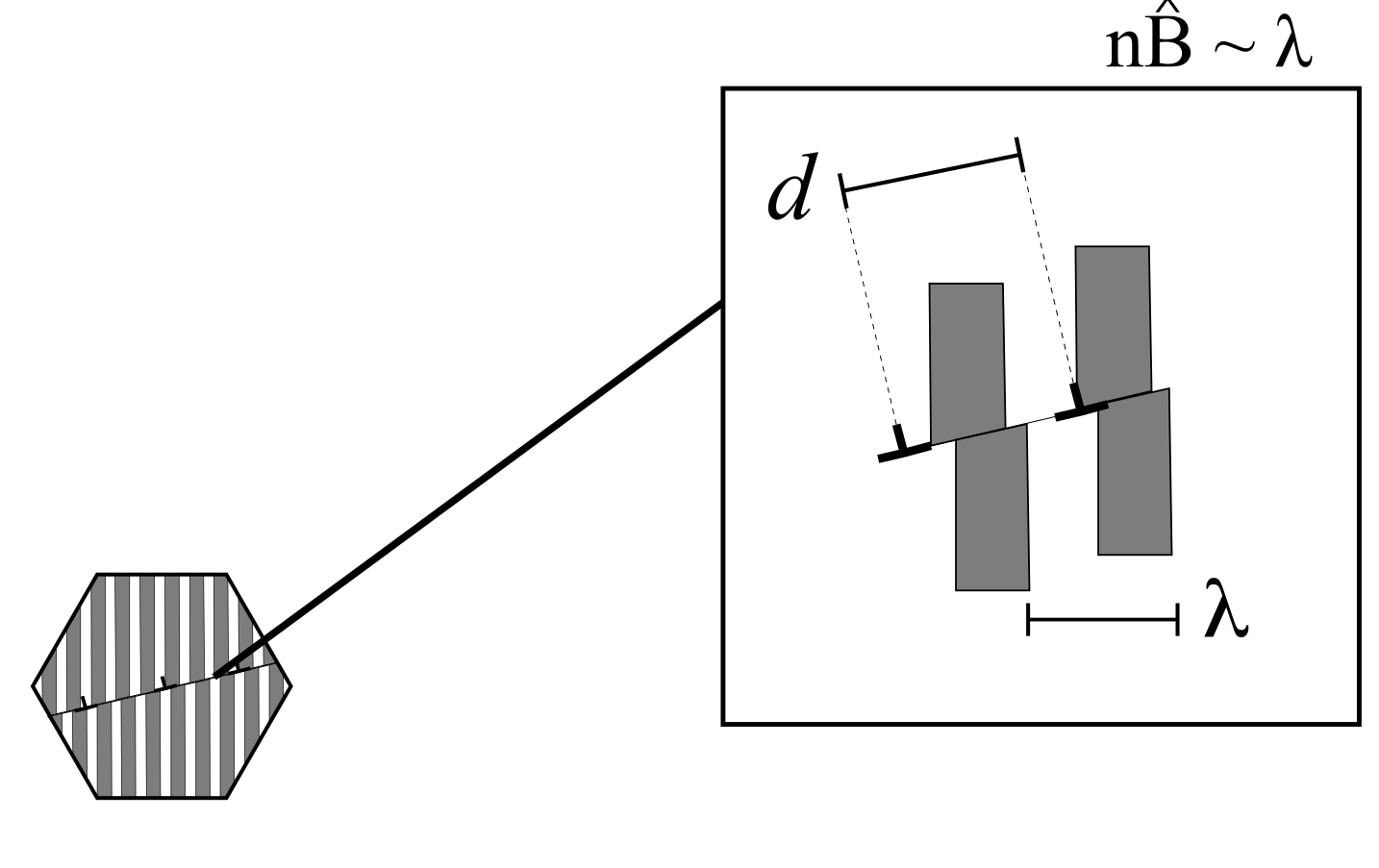}}
\caption{Schematic illustrating the effect of grain refinement on the deformation of the spinodal structure.\label{fig-work-hardening}}
	\end{center}
\end{figure*}

 This is perhaps best illustrated by an example: 
 In a coarse grained material (Fig~\ref{fig-work-hardening}) generating a plastic strain in the range of 1-2\% requires the passage of literally hundreds of dislocations. As a rough estimate, for a 10$\mu$m grain, a 1\% strain equates to 100\,nm. With a Burgers vector for $a/2 \langle 110 \rangle$ or 0.26\,nm, that would require approximately 400 dislocations per grain. Given that the modulation wavelength is of the order of 1-2\,nm, it is easy to conceive that this amount of deformation will locally disrupt the modulated structure, establishing soft pathways for further deformation, especially if concentrated in a handful of slip planes. 

In contrast, if the grain size is reduced to 100\,nm (Fig~\ref{fig-work-hardening-2}), the same 1\% strain would be accommodated by a mere 4 dislocations, giving rise to a total shear displacement which is of a similar magnitude to the modulation wavelength (1--2\,nm, see Figs.~\ref{fig-tem-hrtem-2},~\ref{fig-tem-hrtem-3}. Even if concentrated on a single slip plane, this would not be sufficient to completely disrupt composition modulations with a wavelength of roughly 2\,nm. Furthermore even at slightly higher strains of perhaps 5\%, the mutual repulsion between dislocations would restrict the number of dislocations on closely-aligned slip planes. 

The NC alloy also displays good thermal stability. The grain size of the HPT sample showed no significant increase during brief ageing at 646\,K (a homologous temperature, $T/T_m$, of 0.48) and the hardness remained stable indefinitely at 473\,K ($T/T_m\sim$0.35) (Fig~\ref{fig-vhn}b). 
This is in sharp contrast to pure copper which can coarsen even at room temperature ($T/T_m$=0.22 \cite{GertsmanBirringer1994}). Grain boundary precipitates are scarce even in the peak-aged condition and it seems improbable that the few particles observed would be sufficient to prevent coarsening. However, \textit{ab-initio} calculations indicates that titanium has with -0.504 eV\footnote{The calculation was performed for a $\Sigma$5 $[100]$ $\{012\}$  grain boundary} a moderately strong grain boundary segregation energy \cite{scheiber2020}, and this thermodynamic stabilisation could explain the observed stability.

The successful use of a modulated structure to stabilise plastic flow in a NC alloy despite a strength increase suggests several avenues worthy of further investigation. Quantitative modelling of the different contributors to the strength, particularly grain size and the spinodal structure would assist in optimising the processing route to provide the best combination of strength and ductility.

\section{Conclusions}

Plastic flow in a NC Cu-1.7\,at.\%Ti alloy has been stabilised by introducing a modulated, spinodal structure. 

\begin{itemize}
    \item The hardness ($254\pm2$\,$H_V$) and yield strength (800\,MPa) of the UFG alloy are comparable with micro-grained Cu-Ti alloys containing 3\,at.\%Ti.
	\item The resulting alloy possess much higher ductility than typical UFG alloys, as evidenced by the good elongation to failure of $\sim$9\,\%, area reduction of $\sim$40\% and the presence of microdimples on the fracture surfaces, which are huge compared to the grain size.
	\item The spinodal structure can be retained during post-HPT annealing, leading to increased strength (890\,MPa) while retaining acceptable elongation to failure (7\%).
	\item The improved ductility is attributed to work-hardening due to the presence of the spinodal structure and their increased amplitude.
	\item The deformation response after ageing is not indicative of hardening by annealing.
	\item The success of this approach suggests that spinodal decomposition offers a means to overcome plastic instability in subsequent poor ductility in UFG and NC alloys. This especially holds true in case of heat treatments, where stable solid solutions undergoing anneal hardening would completely lose ductility, a phenomenon not observed here.
\end{itemize}

\section*{Acknowledgements}
The authors would like to thank Peter Kutle\v{s}a for assistance with the HPT experiments, Dr. Sergey Ketov for the use of the arc-melting facility and Silke Modritsch and Gerard for assistance with the SEM and TEM sample preparation.

\bibliography{CuTi}


\end{document}